\documentclass[prb,twocolumn,amsmath,amssymb,superscriptaddress]{revtex4-2}

\usepackage[bookmarks=false]{hyperref}

\usepackage[T1]{fontenc}
\usepackage[applemac]{inputenc}
\usepackage[english]{babel}
\usepackage{graphics}
\usepackage{amssymb}
\usepackage{amsmath}
\usepackage{overpic}
\usepackage{color}
\usepackage[normalem]{ulem}

\renewcommand*{\thefootnote}{\fnsymbol{footnote}}
\newcommand{\be}{\begin{equation}}
\newcommand{\ee}{\end{equation}}
\newcommand{\bea}{\begin{eqnarray}}
\newcommand{\eea}{\end{eqnarray}}

\def\a{\alpha}

\def\d{\delta}

\def\m{\mu}
\def\l{\lambda}

\def\tt{\hat \tau}

\def\s{\sigma}

\def\nn{\nonumber}
\def\lb{\label}
\def\pref#1{(\ref{#1})}

\begin{document}

\author{T. Cea}
\email[]{Co-first author.}
\affiliation{IMDEA Nanoscience, C/ Faraday 9, 28049 Madrid, Spain}
\author{M. Ruiz-Garc\'ia}
\email[]{Co-first author.}
\affiliation{Department of Physics and Astronomy, University of Pennsylvania, Philadelphia, PA 19104, USA}
\author{L. L. Bonilla}
\affiliation{G. Mill\'an Institute, Fluid Dynamics, Nanoscience and Industrial Mathematics and Department of Mathematics, Universidad Carlos III de Madrid, 28911 Legan\'es, Spain}
\author{F. Guinea}
\affiliation{IMDEA Nanoscience, C/ Faraday 9, 28049 Madrid, Spain}
\affiliation{Donostia International Physics Center,
Paseo Manuel de Lardizabal 4. 20008 San Sebastian, Spain}
\date{\today}

\title{
A numerical study of the rippling instability driven by electron-phonon coupling in graphene
}


\begin{abstract}
Suspended graphene exhibits ripples of size ranging from 50 to 100 {\AA} and height $\sim$10{\AA}, however, their origin remains undetermined. Previous theoretical works have proposed that rippling in graphene might be generated by the coupling between the bending modes and the density of electrons. These theoretical studies proposed that, in the thermodynamic limit, a membrane of single layer graphene becomes unstable for large enough electron-phonon coupling, which signals a phase transition from a flat phase to a rippled one. Here, we find the stable configuration of a suspended monolayer of graphene at $T=0$ by minimizing the average energy of a membrane where the Dirac electrons of graphene couple to elastic classical deformation fields. We find that the electron-phonon coupling controls a transition from a stable flat configuration to a stable rippled phase. We propose a scaling procedure that allows us to effectively reach larger system sizes. We find that the critical value of the coupling, $g_c$, rapidly decays as the system increases its size, in agreement with the experimental observation of an unavoidable stable rippled state for suspended graphene membranes. This decay turns out to be controlled by a power law with a critical exponent  $\sim 1/2$. Consistent arguments based on bifurcation theory indicate that the phase transition is discontinuous at large scaling parameter $k$, that the jump in the order parameter decreases as $k^{-1/2}$, and that the phase transition becomes continuous at $k=\infty$, with the order parameter scaling as $(g-g_{c,\infty})^{1/4}$.

\end{abstract}
\maketitle

Graphene is a well studied material ~\cite{Novoselov_science04,Novoselov_nat05,Novoselov_PNAS05,Zhang_nat05}. It has attracted the attention of a broad scientific community due to its unconventional electronic properties~\cite{Novoselov_nat05,castroneto_RevModPhys09,Zhang_nat05,Novoselov_nat05,morozov_prl08,chen_nat08} and its exceptional mechanical properties~\cite{Lee_science08}. Its one-atom thickness makes it the perfect candidate to test the effect that thermal fluctuations can have on its elastic properties as a crystal membrane~\cite{nelson_book,nelson_JphysF1987,ledoussal_prl92,Letal15,Betal15,kovsmrlj2016response,bowick2017non,LR18,morsheda2019buckling}. Transmission electron microscopy shows that suspended graphene membranes exhibit stable ripples, remarkably different from the thermal fluctuations arising on a  flat configuration~\cite{meyer_nat07}. The out-of-plane deformations of free-standing graphene influence its electronic properties, thereby changing the electrical conductivity~\cite{mariani_prl08,castro_prl10,mariani_prb10} and generating spatially varying gauge potentials~\cite{guinea_prb08,guinea_SSC09,castroneto_RevModPhys09}. The latter  induce charge inhomogeneity~\cite{kim_EPL08} and underlie the formation of electron-hole puddles~\cite{gibertini_prb12}.

Previous works of rippling and buckling phenomena have studied a simplified model of Ising spins (modelling electronic degrees of freedom) coupled to an elastic membrane. This model exhibits a rich phase diagram with flat, buckled and rippled phases \cite{BonillaJSM_2012,ruiz2015ripples,ruiz2016stm}. In $1$D, the model can be analytically solved and shown to have first and second order phase transitions when the temperature and the interaction between the spins are controlled \cite{ruiz2017bifurcation}. However, a more realistic approach explaining the origin of rippling in suspended graphene and involving electronic degrees of freedom is not yet developed \cite{fasolino_nat07}. Previous theoretical studies~\cite{gazit_prb09,sanjose_prl11,guinea_prb14,bonilla2016critical,PhysRevE.80.041117,PhysRevB.92.155428,GDMP16,SGKB20,FLK07,KL14,GKM17,G14} proposed that the coupling between elastic and electronic degrees of freedom might be at the origin of rippling in graphene, which would arise as a phase transition controlled by the coupling strength. However, in these works, the appearance of ripples is inferred indirectly through a vanishing renormalized bending modulus~\cite{sanjose_prl11} or by a postulated soft mode at finite momentum~\cite{guinea_prb14}. A direct numerical calculation of ripples, and its effect on the structure of the electronic band, is still missing and motivates our work. Here we use a realistic model of suspended graphene membranes coupled to their electronic degrees of freedom. Through numerical simulations, we show that stable ripples in suspended graphene membranes can spontaneously arise as a phase transition from a flat state as the electron-strain coupling increases. We also show that as the coupling increases a gap opens in the band structure. Finally, the critical value of the coupling parameter tends to zero as the system size increases, which agrees with the experimental observation of unavoidable rippling of suspended graphene membranes.

In our model, we consider a {\em classical} elastic membrane with periodic boundary conditions, coupled to the {\em quantum} Hamiltonian for Dirac electrons in graphene and discretized in an effective hexagonal lattice, at zero temperature. The spatial distributions of strain, heights and electronic density are strongly correlated with each other in the rippled phase. Thus, ripples are not triggered by a buckling instability of a clamped or supported finite membrane under tension. Instead, we show that they arise on a large sheet when the electron-strain coupling is large enough to generate stable rippled configurations, characterized by non-homogeneous spatial distributions of strain. We can effectively simulate membranes of larger size by properly defining a scaling parameter $k=a/a_0$, where $a$ is the effective lattice constant and $a_0=2.46${\AA} is that of pristine graphene. In the large system limit, the transition happens at very low coupling, which agrees with experimental observations of stable corrugated membranes~\cite{meyer_nat07}. Even more interesting, we find that the critical value of the electron-strain coupling decreases to zero as a power law $\sim k^{-1/2}$ as the scaling size $k$ increases.

The structure of this paper is the following: we introduce the physical model in section \ref{sec_model}; the numerical scheme used to solve the model is presented in section \ref{sec_it}; section \ref{sec_sca} carries out a scaling analysis that allows us to study samples of larger size; and finally, we discuss our results in section \ref{sec_dis}.
 
\section{the model}
\label{sec_model}
We describe the elastic deformations of a membrane of monolayer graphene by the vector field $\mathbf{u}=(u_x,u_y,h)$, where $u_x,u_y$ are the in-plane deformations of the membrane with respect to the equilibrium position and $h$ the out-of-plane shift. In the long-wavelength limit, the elastic energy of the membrane can be defined in terms of the strain tensor $u_{ij}=\frac{1}{2}\left(\partial_iu_j+\partial_ju_i+\partial_ih\partial_jh\right)$ as \cite{landau_book}:
\bea
E_{el}=\frac{1}{2}\int\,d^2\mathbf{r}\left[\kappa\,(\nabla^2 h)^2+\lambda u_{ii}^2+2\mu u_{ij}^2\right], \label{Hel}
\eea
where repeated indices $i,j=x,y$ are implicitly summed. Here $\kappa$ is the bending rigidity and $\l,\m$ are the Lam\'e coefficients. The energy \eqref{Hel} and the displacement vector are discretized on the honeycomb lattice. We assume that there is a direct coupling between the electronic charge density and the strain, and that the kinetic energy of the electrons is described by the nearest neighbors tight binding approximation. Upon discretizing the Dirac Hamiltonian for electrons in graphene, we have the following Hamiltonian:
\begin{align}\label{H_e}
\hat{H}_e&=-t\sum_{\langle \mathbf{R},\mathbf{R'} \rangle \sigma}\left(\hat{a}^{\dagger}_{\mathbf{R}\sigma}\hat{b}_{\mathbf{R'}\sigma}+h.c. \right) \\
&-g\sum_\mathbf{R}(\hat{n}(\mathbf{R}) - n_0)u_{ii}(\mathbf{R}).
\label{H_e}
\end{align}
Here $\hbar=1$, $t$ is the hopping integral which, for pristine graphene, is approximatively $t_0=2.7$~eV, and $\langle \mathbf{R},\mathbf{R'} \rangle$ indicate nearest neighbors sites in the hexagonal lattice. $\hat{a}$ and $\hat{b}$ are the annihilation operators for a fermion in the $A$ and $B$ sublattices of the honeycomb lattice, respectively, $\sigma$ is the spin index and $g$ is the coupling strength. In \eqref{H_e}, we consider a scalar electron-strain potential that couples the charge density at each lattice point with the strain at that point of the membrane. Other coupling terms considered in previous works, such as a gauge field that induces current fluctuations, see e.g. \cite{vozmediano_physrep2010}, are suppressed in the long-wavelength limit \cite{sanjose_prl11,guinea_prb14}. In \eqref{H_e}, $n_0=1$ is the equilibrium occupation number for the undoped system, and $\hat{n}(\mathbf{R})$ is the local occupation number:
\bea
\hat{n}(\mathbf{R})=
\begin{cases}
\sum_\s \hat{a}^{\dagger}_{\mathbf{R}\sigma}\hat{a}_{\mathbf{R}\sigma}\text{, if $\mathbf{R}\in$ $A$} \\
\sum_\s \hat{b}^{\dagger}_{\mathbf{R}\sigma}\hat{b}_{\mathbf{R}\sigma}\text{, if $\mathbf{R}\in$ $B$}
\end{cases}
\eea
Within this approximation, the graphene Fermi velocity is $v_F= \sqrt{3} t_0 a_0/2\hbar$.
The Dirac dispersion is an approximation to the electronic structure of graphene at low energies, and it is, in turn, based on an atomistic model which considers the two bands formed by the $p_z$ carbon orbitals.
Note that the lattice discretization of the Dirac equation that we use, Eq. \pref{H_e}, although based on sites in a honeycomb lattice, is not directly related to atomic orbitals.
 Finally, the complete Hamiltonian for the graphene membrane coupled to the electron density distribution is given by:
\bea\lb{Htot}
\hat{H}=E_{el}+\hat{H}_e.
\eea
Typical values for graphene are $\kappa=0.82$~eV, $\lambda=19.67$~eV$/a_0^2$ and $\mu=57.13$~eV$/a_0^2$, where $a_0=2.46${\AA} is the lattice constant. Finally, we have: $\lambda_0=3.25$~eV{\AA}$^{-2}$ and $\mu_0=9.44$~eV{\AA}$^{-2}$~\cite{fasolino_nat07,zakharchenko_prl09}. There is no consensus about the magnitude of the electron-phonon coupling $g$; recent estimates give values in the range $g\sim 4-50$~eV~\cite{ono_jpsj1966,suzuura_prb02,choi_prb10}.

It is worth noting that we do not include the Coulomb repulsion between the electrons.
A discussion of the role of Coulomb interactions can be found in Ref. \cite{gazit_prb09}.
The deformation of the membrane generates charge ''puddles''
that have been shown to have a small quantum capacitance, and hence would not prevent the rippled phase \cite{gazit_prb09,sanjose_prl11}.
The overall effect of the Coulomb interaction is to increase the critical value of the electron-phonon coupling
as compared to its bare value \cite{guinea_prb14}, and would not change our main results.
In addition, the long range part of the Coulomb interaction is screened by the environment in many realistic setups, making its effect negligible.

We study the Hamiltonian of Eq. \pref{Htot} within the Born-Oppenheimer (adiabatic) approximation, in which the quantum problem for the electrons is solved by treating the elastic fields classically. In this approach, the displacements $\mathbf{u}$ enter in the quantum problem
as external parameters, via the interaction term proportional to $g$ in \eqref{H_e}. The total energy of the membrane is then a functional of the displacement fields:
\bea\lb{Etot}
E_{tot}[\mathbf{u}]=\left\langle \hat{H}  \right\rangle,
\eea
where the brackets denote the quantum average as computed by means of $\hat{H}$ itself:
\be\label{mean}
\left\langle\hat{O}\right\rangle\equiv\frac{\mathrm{Tr}\left\{ e^{-\hat{H}/K_BT} \hat{O}\right\}}{\mathrm{Tr}\left\{ e^{-\hat{H}/K_BT} \right\}},
\ee
where $K_B$ is the Boltzmann constant, $T$ the temperature and $\hat{O}$ any operator.

In this work we study the equilibrium configurations of the membrane at $T=0$, which reduces the quantum averages in \eqref{Etot} to sums over the eigenvalues of the quantum operators. At $T=0$, the equilibrium state of the system minimizes the total energy \pref{Etot} with respect to the displacements. To solve the minimization problem, we consider its Euler-Lagrange equations. They are the eigenvalue problem for the Hamiltonian \eqref{Htot} and the extremal equations (equilibrium condition):
\bea\lb{eq_cond}
\frac{\delta E_{tot}[\mathbf{u}]}{\delta \mathbf{u}(\mathbf{r})}=0.
\eea
Thus, the Euler-Lagrange equations constitute a nonlinear eigenvalue problem because the solution of \eqref{eq_cond} enters the Hamiltonian $\hat{H}$ through \eqref{H_e}, which, in turn, modifies \eqref{eq_cond}. At $T=0$ the functional derivatives of $E_{tot}$ can be performed by means of the Feynman-Hellmann theorem, which allows to switch the order in which the derivatives and the quantum average are computed:
$\frac{\delta E_{tot}[\mathbf{u}]}{\delta \mathbf{u}(\mathbf{r})}=  \left\langle \frac{\delta \hat{H}}{\delta \mathbf{u}(\mathbf{r})}  \right\rangle$. Then \pref{eq_cond} becomes:
\bea
\frac{\delta E_{el}[\mathbf{u}]}{\delta \mathbf{u}(\mathbf{r})} - g\,\sum_{\mathbf{R}}[\langle \hat{n}(\mathbf{R})\rangle-n_0]\frac{\delta u_{ii}(\mathbf{R})}{\delta \mathbf{u}(\mathbf{r})} =0,
\label{spe}
\eea
which, in turn, is equivalent to the following system:
\begin{widetext}
\begin{subequations}\label{eq_cond2}
\bea
\lambda \partial_x\left(\partial_xu_x+\partial_y u_y+\frac{\left|\bm{\nabla} h\right|^2}{2}\right)+
\mu \partial_x\left[	2\partial_xu_x+\left(\partial_xh\right)^2	\right]&+&
\mu\partial_y\left(\partial_yu_x+\partial_xu_y+\partial_xh\partial_yh		\right)= g\, \partial_x\delta\rho(x,y), \lb{uxeq}\\
\lambda \partial_y\left(\partial_xu_x+\partial_y u_y+\frac{\left|\bm{\nabla} h\right|^2}{2}\right)+
\mu \partial_y\left[	2\partial_yu_y+\left(\partial_yh\right)^2	\right]&+&
\mu\partial_x\left(\partial_yu_x+\partial_xu_y+\partial_xh\partial_yh		\right)= g\, \partial_y\delta\rho(x,y), \lb{uyeq}\\
\lambda\bm{\nabla}\cdot\left[\left(\partial_xu_x+\partial_yu_y+\frac{\left|\bm{\nabla}h\right|^2}{2}\right)\bm{\nabla}h\right]+
\mu
\partial_x\left[2\partial_xu_x\partial_xh\right.&+&\left.\left(\partial_yu_x+\partial_xu_y\right) \partial_yh+ \left|\bm{\nabla}h\right|^2\partial_xh \right]+\nn\\
+
\partial_y\left[2\partial_yu_y\partial_yh+\left(\partial_yu_x+\partial_xu_y\right) \partial_xh+ \left|\bm{\nabla}h\right|^2\partial_yh \right]
&-&\kappa \left(\nabla^2\right)^2h=g\bm{\nabla}\cdot\left(	\delta\rho\bm{\nabla}h	\right)\!.\lb{heq}
\eea
\end{subequations}
\end{widetext}
We show here equations \eqref{eq_cond2} in the continuous limit for convenience, but we will use their discrete counterpart for the numerical calculations, see appendix \ref{app_2} for more details.
In \eqref{eq_cond2}, $\delta\rho(\mathbf{r})=\langle\hat{n}(\mathbf{r})\rangle-n_0$  is the charge distribution corresponding to the ground state of the electronic Hamiltonian: $\hat{H}_e$, which also depends on the displacements $\mathbf{u}$. Recapitulating, to find the minimum of \eqref{Etot}, we have to solve equations \eqref{eq_cond2} for $\mathbf{u}$ while simultaneously diagonalizing the electronic Hamiltonian (which also depends on $\mathbf{u}$). This is done by an iterative process, as explained in the next section. 

The physical role played by each term in Eqs.~\pref{eq_cond2} can be made more clear if we rewrite the equations in terms of an Airy potential, $\chi$, which is defined as:
$
\partial^2_x\chi=\lambda u_{ii}+2\mu u_{yy}-g\delta\rho$, 
$\partial^2_y\chi=\lambda u_{ii}+2\mu u_{xx}-g\delta\rho$, 
$\partial_x\partial_y\chi=-2\mu u_{xy}$.
As shown in appendix \ref{app_1}, Eqs.~\pref{eq_cond2} are equivalent to the F\"{o}ppl-von K\'arm\'an equations \cite{foppl_book,karman_book}:
\begin{subequations}\lb{eq_cond3}
\bea
\kappa  \nabla^2 h-2[\chi,h]&=&0\\
\frac{1}{Y}\nabla^2\chi+[h,h]&=&-\frac{g}{2B}\nabla^2\delta \rho,\lb{curv_eq}
\eea 
\end{subequations}
where $Y=\frac{4\mu(\l+\m)}{\l+2\m}$ and $B=\l+\m$
are the Young and compression moduli, respectively. Also, $[\chi,h]\equiv \frac{1}{2}\left[  \partial^2_x \chi\partial^2_y h+  \partial^2_y \chi\partial^2_x h-2\left(\partial_x\partial_y\chi\right)\left(\partial_x\partial_yh\right) \right]$,
so that $[h,h]$ is the curvature of the membrane, as defined in Ref.~\cite{landau_book}. Thus, the spatially varying electronic density, which depends on the eigenvectors of $\hat{H}_e$, acts as the source of curvature according to Eq.~\pref{curv_eq}. In turn, the curved membrane modifies the potential term in the Hamiltonian that is being diagonalized ($\hat{H}_e$).

\section{Self-consistent numerical approach to solve the coupled problem }

\label{sec_it}

To solve the system of equations \eqref{eq_cond2} coupled to the eigenvalue problem for the electrons we use all the expressions discretized  on the hexagonal lattice with coordinates $\mathbf{R}$. For more details see appendix \ref{app_2}. The contribution of the electrons to the total energy is given by the Hamiltonian \eqref{H_e}, which we can rewrite as,
\begin{align}
\hat{H}_e= \hat{H}_e^0 + \hat{H}_e^1
\label{H_e_n}
\end{align}
where,
\begin{align}
\hat{H}_e^0 =
-t\sum_{\langle \mathbf{R},\mathbf{R'} \rangle \sigma}\left(\hat{a}^{\dagger}_{\mathbf{R}\sigma}\hat{b}_{\mathbf{R'}\sigma}+h.c. \right) -g\sum_\mathbf{R}\hat{n}(\mathbf{R}) u_{ii}(\mathbf{R}),
\label{H_e0}
\end{align}
and,
\begin{equation}
 \hat{H}_e^1=g n_0 \sum_\mathbf{R} u_{ii}(\mathbf{R}).
\label{H_e1}
\end{equation}

$ \hat{H}_e^1$ only depends on the strain and can be treated as a contribution to the elastic energy. $\hat{H}_e^0$ is the contribution to the energy that depends on the charge distribution. In fact, for a system of $N$ atoms, this term is numerically computed using a tight-biding matrix with $N\times N$ elements. Element $ij$ will be $-t$ if atoms $i$ and $j$ are neighbors and $0$ otherwise. Finally, the diagonal elements of this matrix take on values $-g u_{ii}(\mathbf{R})$ (where $\mathbf{R}$ stands for the position in the lattice corresponding to that element of the matrix).

To solve the nonlinear eigenvalue problem with the discretized version of equations \eqref{eq_cond2}, we use an iterative procedure. Given an initial condition $\mathbf{u}_0$, we first compute the local strain $u_{ii}(\mathbf{R})$ and consequently diagonalize the electronic Hamiltonian \eqref{H_e0}. We consider the problem without doping, where the electronic distribution corresponds to the half-filled energy band, taking into account the spin degeneracy. Let  $E_1\le E_2\le \dots \le E_{N}$ be the eigenvalues of \eqref{H_e0}, and $U_\a(\mathbf{R})$ their corresponding eigenvectors. At $T=0$, the local occupation number $n(\mathbf{R})=\langle\hat{n}(\mathbf{R})\rangle$  is given by:
\bea
n(\mathbf{R})=2 \sum_{\a\le N/2} \left| U_\a(\mathbf{R}) \right|^2,
\eea
where the factor of two accounts for the spin degeneracy and the sum over the lowest half of the spectrum sets the Fermi level at $E_F=E_{N/2}$ if $E_{N/2}$ is a simple eigenvalue (otherwise the rule needs to be modified in an obvious way to ensure that Pauli's principle holds). The resulting occupation number, $n(\mathbf{R})$, is then used as a new input for the elasticity equations \pref{eq_cond2} (see appendix \ref{app_2} for discretized version). The procedure is iterated until the displacement fields, $\mathbf{u}$, converge.

Once a solution is found,
the total energy of the membrane is computed as the sum of the elastic and electronic contributions:
$E_{tot}=E_{el}+E_e$,
where $E_{el}$ is given by the discretized version of Eqs. \eqref{Hel} and $E_e$ is equal to $2\sum_{\a\le N/2} E_\a $ plus the discretized version of \eqref{H_e1}. Note that a completely flat configuration with an homogeneous distribution of charge is always a solution of the Eqs. \eqref{eq_cond2}, although it may not be stable. 

\begin{figure}
 
\includegraphics[scale=0.4]{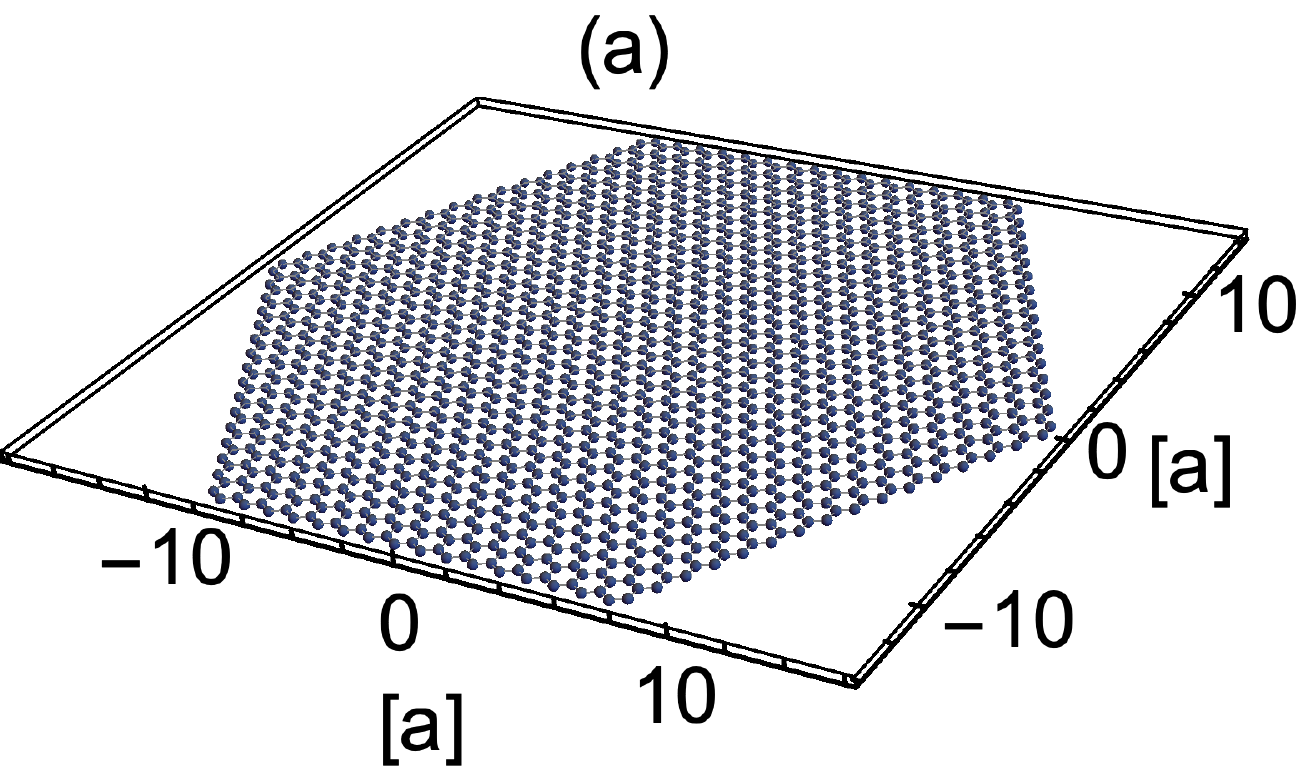}

\includegraphics[scale=0.4]{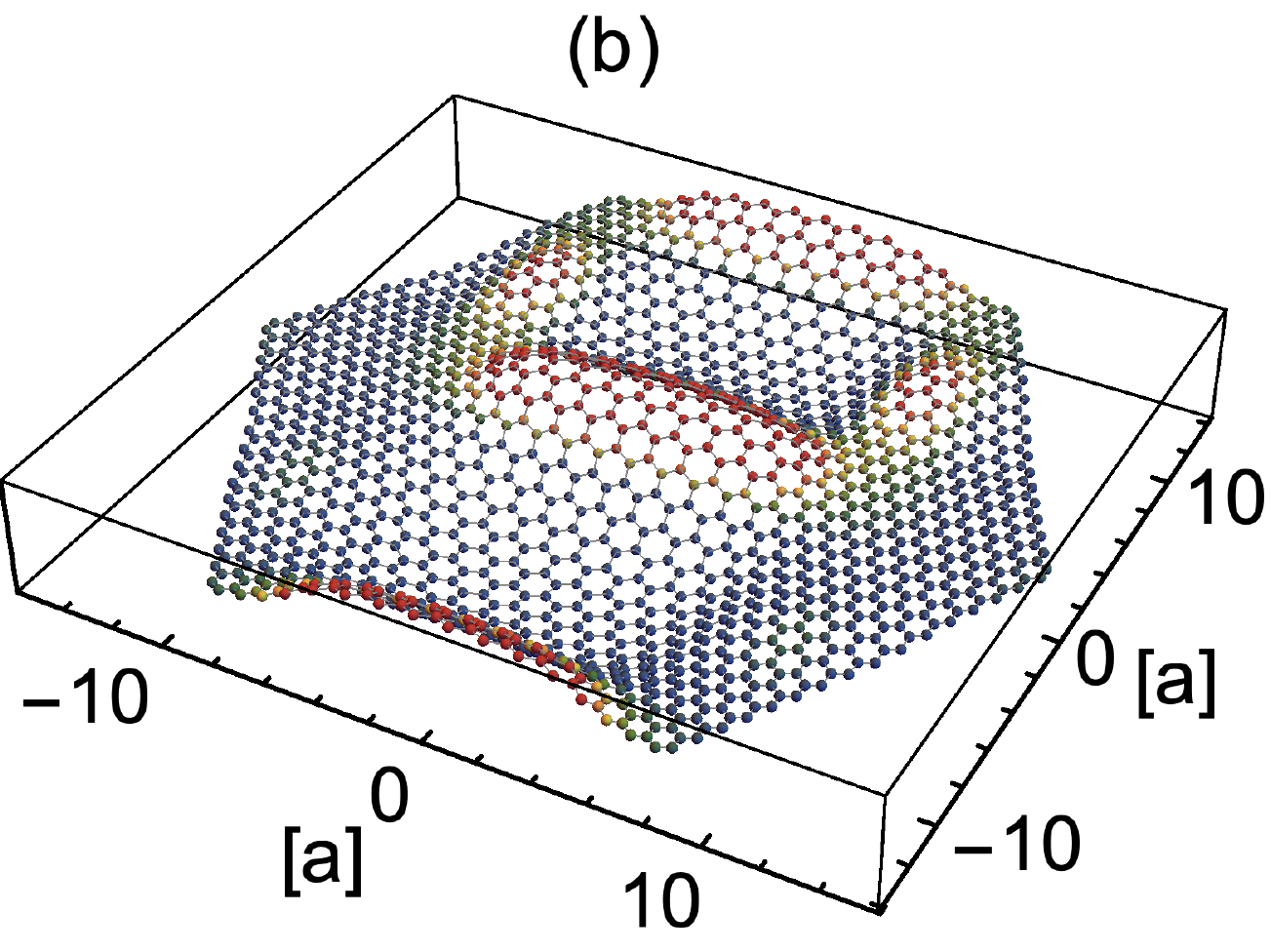}

\includegraphics[scale=0.4]{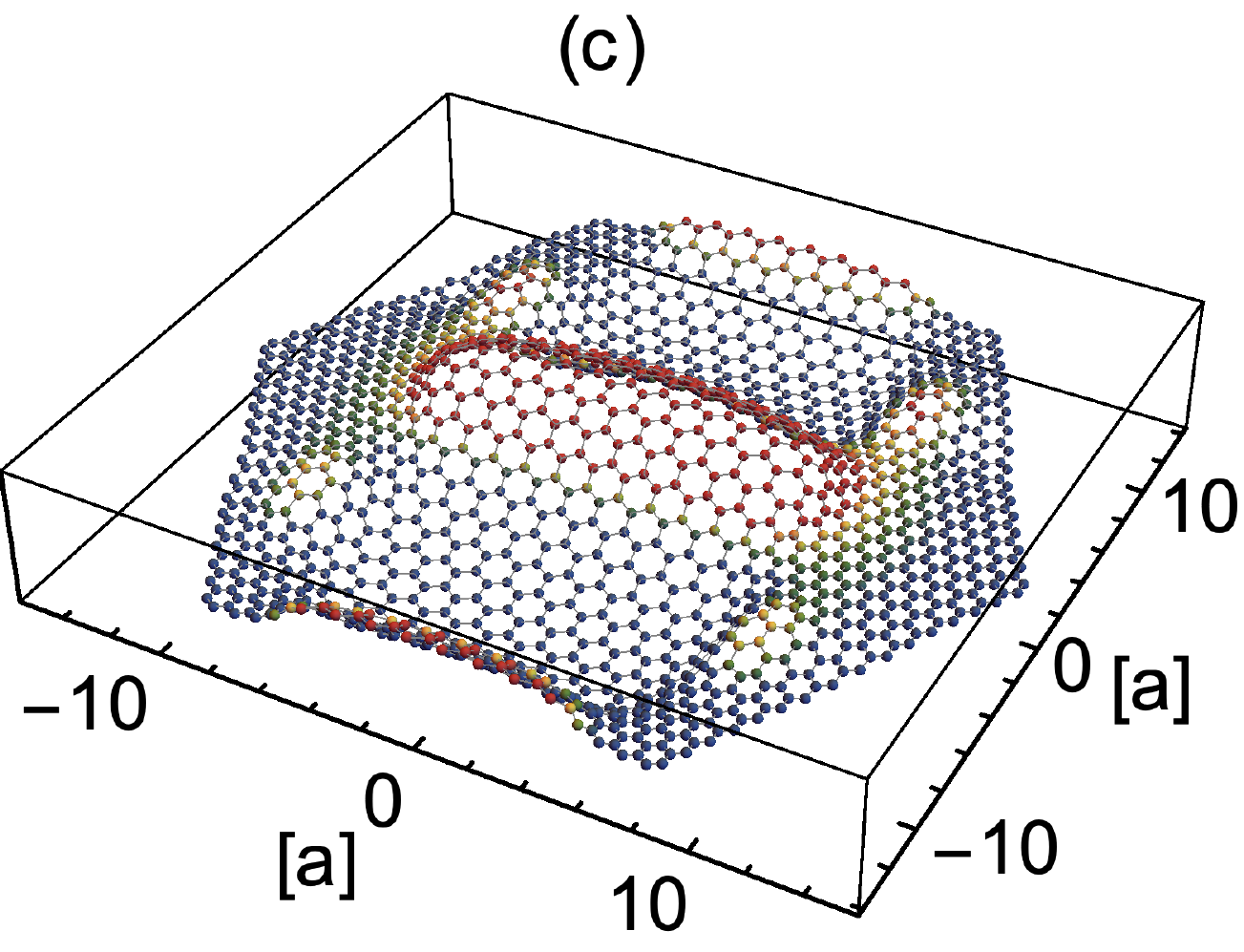}

\caption{
Solution of the iterative scheme for three different values of $g$. Panels (a)-(c) correspond to $g=8t, 10.5t$ and $12t$, respectively. For this system size the flat solution is stable for $g=8t$, when we increase $g$ a rippled phase appears (panels (b) and (c)).
}
\label{fig_comparig_3_g}
\end{figure}

Figure \ref{fig_comparig_3_g} shows three solutions of equations \eqref{eq_cond2}, obtained within the iterative method described above. We use a hexagonal lattice with $1536$ atoms and periodic boundary conditions. The coupling parameter for these cases takes the value $g=8t_0, \ 10.5t_0, \ 12t_0$. We start the iterative procedure with a Gaussian profile for the height field, $h$, peaked in the center of the membrane. The stationary solutions shown in the figure display a flat configuration for $g=8t_0$ and a non-homogeneous distribution of the height for $g=10t_0$ and $12t_0$. This shows that the coupling introduced in the equations is enough to bring the system out of the flat configuration and to stabilize rippled configurations. This suggests that there is a rippling phase transition controlled by $g$.

Finally, it would not have been practical to use the F\"oppl-von K\'arm\'an equations \eqref{eq_cond3} instead of the elasticity Eqs.~\eqref{eq_cond2}: to recover the stress and strain tensors from the potential $\chi$, we need to differentiate it twice; see Appendix \ref{app_1}. Since we solve spatially discretized equations, the additional differentiations would involve extra approximations when we discretize them.

\section{SCALING ANALYSIS}
\label{sec_sca}

The main limitation of a numerical approach is extrapolating its results to membranes of realistic sizes. Even though small samples of the order of $10^3$ carbon atoms already give interesting results,  our goal is to understand the rippling transition in much bigger membranes that may credibly approximate infinite ones. We aim to study the behavior of the rippling transition at a critical value $g_c$ of the coupling constant for large scales. To circumvent the limitations of simulating a large, but not infinite, amount of atoms, we solve Eqs.~\eqref{spe} varying the scaling parameter $k=a/a_0$ (introduced above), which controls the effective size of the sample. Varying $k$ is equivalent to defining  a new honeycomb lattice that is a renormalized version of the original graphene membrane. In the scaled membrane, each point does not correspond to a single carbon atom but to a coarse grained set of unit cells. We scale the terms of our equations so that the elastic energy is independent of the scaling. Then the Lam\'e coefficients must scale as: $\lambda= \lambda_0/k^2$ and $\mu=\mu_0/k^2$, while the bending energy, $\kappa$, remains constant. Moreover, requiring the Fermi velocity not to vary upon scaling implies that the hopping parameter scales as: $t=t_0/k$.

 \begin{figure}
\includegraphics[scale=0.3]{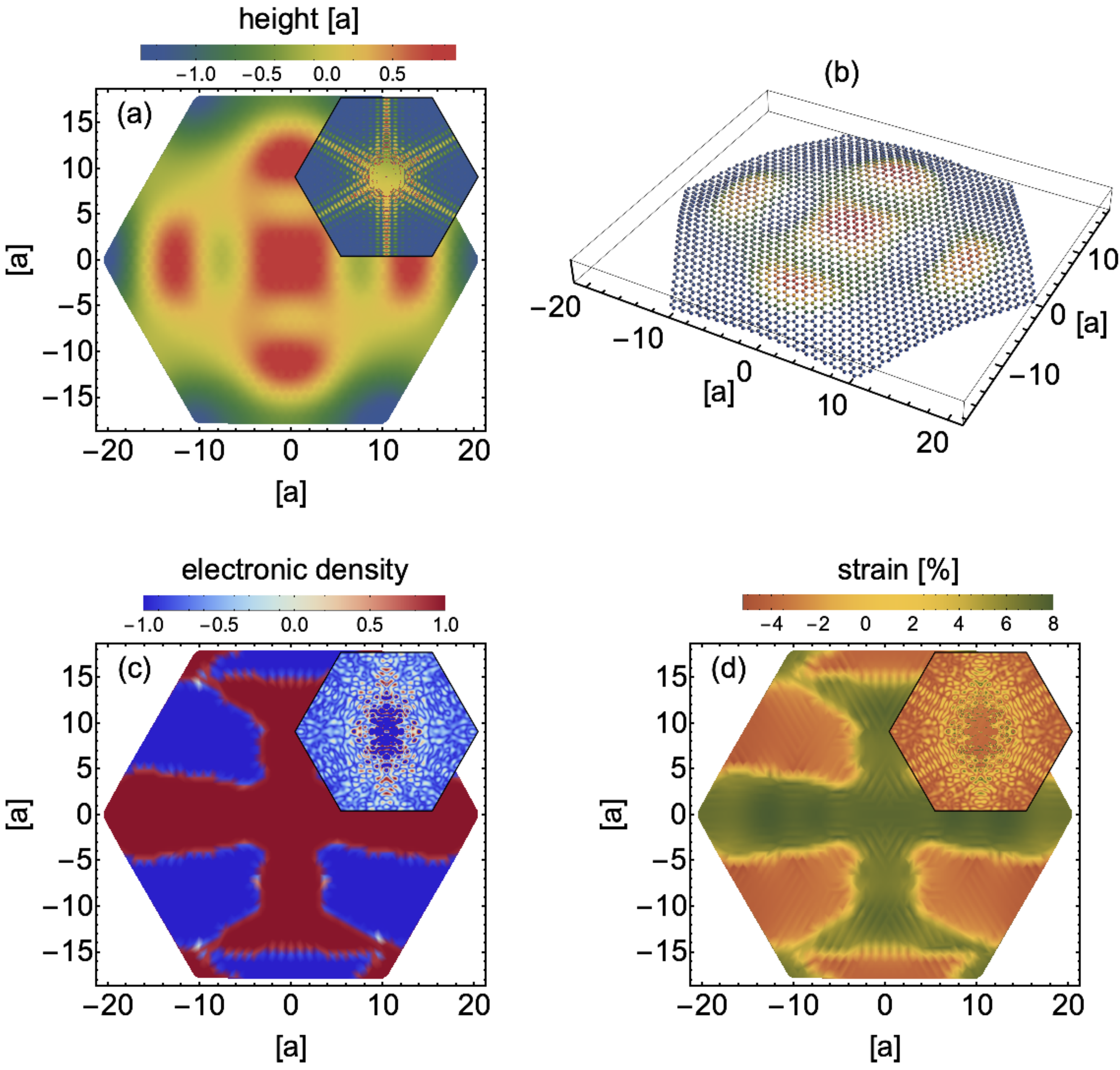}
\caption{
Numerical solution of the elasticity equations coupled to the quantum electronic problem for $g=2t_0$. Here we simulate a system of $N=2646$ lattice nodes with periodic boundary conditions and lattice constant $a=50a_0$. Panel (a)-(b): heights distribution shown as a contour plot and 3D plot, respectively. Panel (c): electronic occupation number counted from half-filling. Panel (d): local strain. The insets show the corresponding Fourier amplitudes in the first Brillouin zone of pristine graphene.
}
\label{fig1}
\end{figure}

\begin{figure}
\includegraphics[scale=0.25]{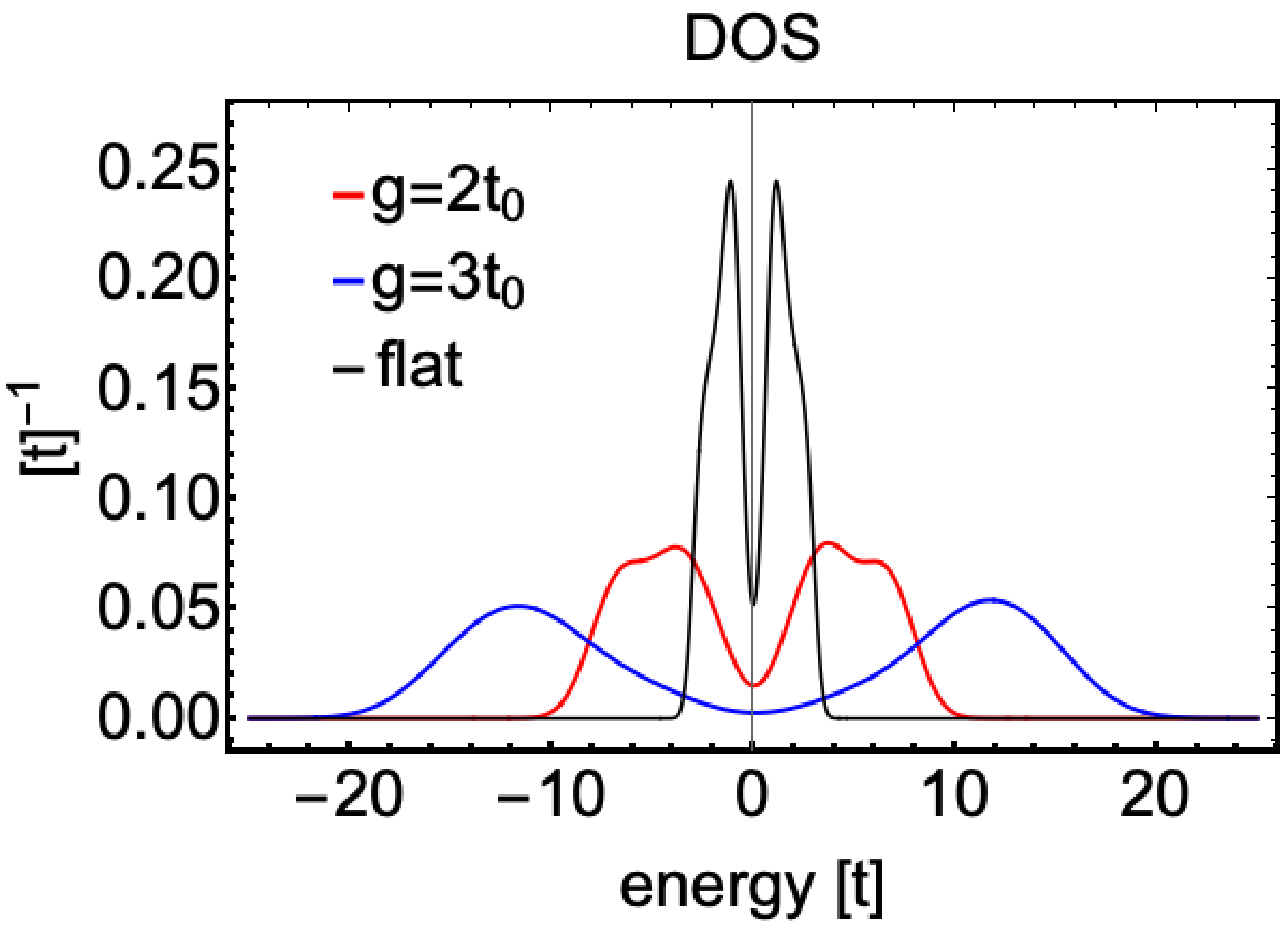}
\caption{Normalized electronic density of the states obtained for different values of the electron-strain coupling $g$. As the coupling increases the rippling is more pronounced, pushing the spectral weight towards higher energies and opening a gap at the Fermi level. The black line, corresponding to the undeformed membrane, represents the typical spectrum of free electrons in graphene in the tight binding approximation. These results correspond to a system of $N=2646$ sites with periodic boundary conditions and lattice constant $a=50a_0$, as in Figure \ref{fig1}.
}
\label{fig2}
\end{figure}

Figure \ref{fig1} shows a solution of the iterative scheme for a membrane with $N=2646$ lattice nodes, periodic boundary conditions, lattice constant $a=50a_0$ and  $g=2t_0$. The scaling parameter is $k=50$. Fig.~\ref{fig1}(a)-(b) depict the space distribution of the deformation fields, Fig.~\ref{fig1}(c) shows the charge distribution, and the insets in the panels show their corresponding Fourier transforms. Note that the flat configuration is already unstable for $g=2$ if $k=50$, whereas it is stable if $k=1$ as in  Fig.~\ref{fig_comparig_3_g}. The values of the strain, shown in Fig.~\ref{fig1}(d), range approximatively from $-5\%$ to $8\%$, in agreement with the experimental observations, and well below the threshold for fracture \cite{meyer_nat07,fasolino_nat07}. Note the correlation of the strain and the charge distribution due to the coupling between the two. To visualize the effect of the electron-strain coupling on the electronic spectrum, in Figure \ref{fig2} we show the electronic density of states for $g/t_0=2$ and $3$, normalized to $1$. The continuum black line represents the Dirac limit corresponding to the flat membrane. As $g$ increases, the spectral weight is pushed to higher energies, up to $\sim 20 t$, while the band edge for free electrons in graphene is $3t$. Furthermore, the data suggests that increasing $g$ may open a gap in the electronic spectrum in the rippled phase.
We note that it has been previously shown in Ref. \cite{choi_prb10} that the electronic spectrum is not gapped, even under very large strains.
However, Ref. \cite{choi_prb10} considers the case of a uniform strain, while our results account for a finite strain gradient.
Although we do not rigorously prove the existence of a gap, the opening of a gap in the Dirac spectrum is an effective way of lowering the electronic energy
and, in the present context, may be linked to the Anderson localization induced by disorder \cite{anderson_58}.

 \begin{figure}
\includegraphics[scale=0.35]{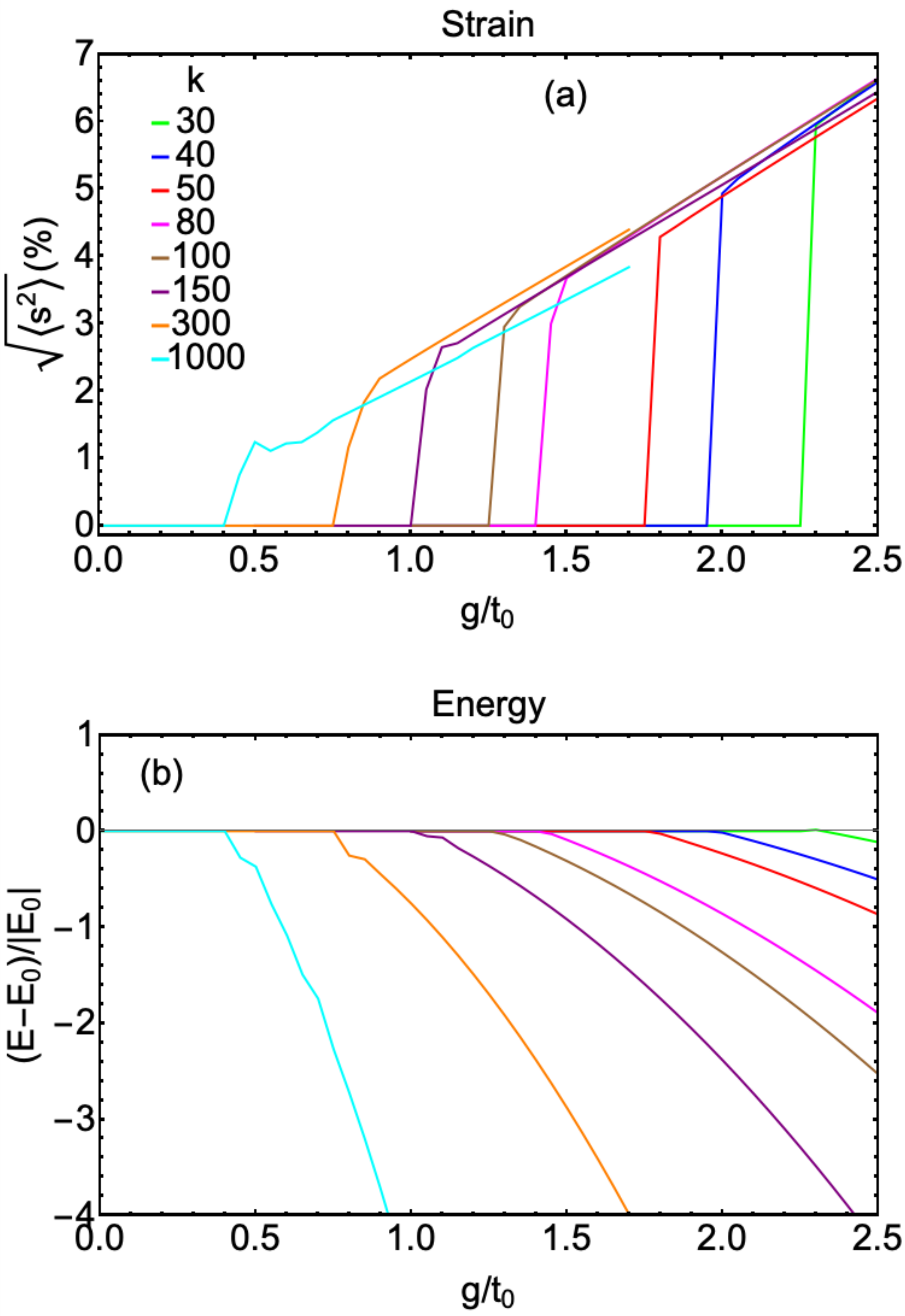}
\includegraphics[scale=0.35]{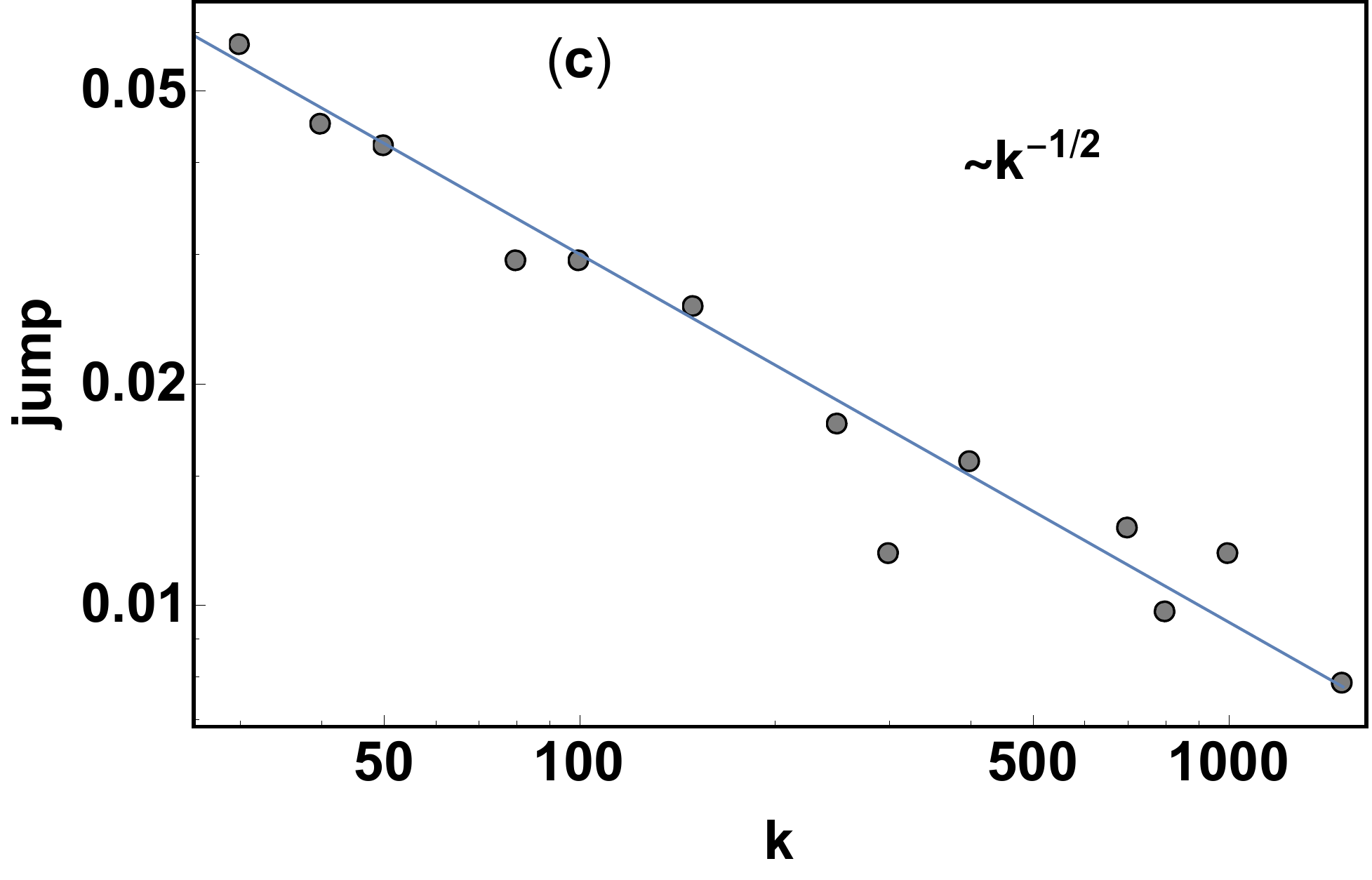}
\caption{Panel (a): geometric average of the local strain
as a function of $g$ for different values of the scaling parameter $k$.
Panel (b): total energy \eqref{Htot} for the same simulations displayed in (a). As $k$ increases the transition occurs for a smaller critical value of $g$. 
Panel (c): size of the jump in the average strain of Panel (a) versus $k$. The line is a fit $\sim k^{-1/2}$.
}
\lb{fig3}
\end{figure}

 \begin{figure}
\includegraphics[scale=0.45]{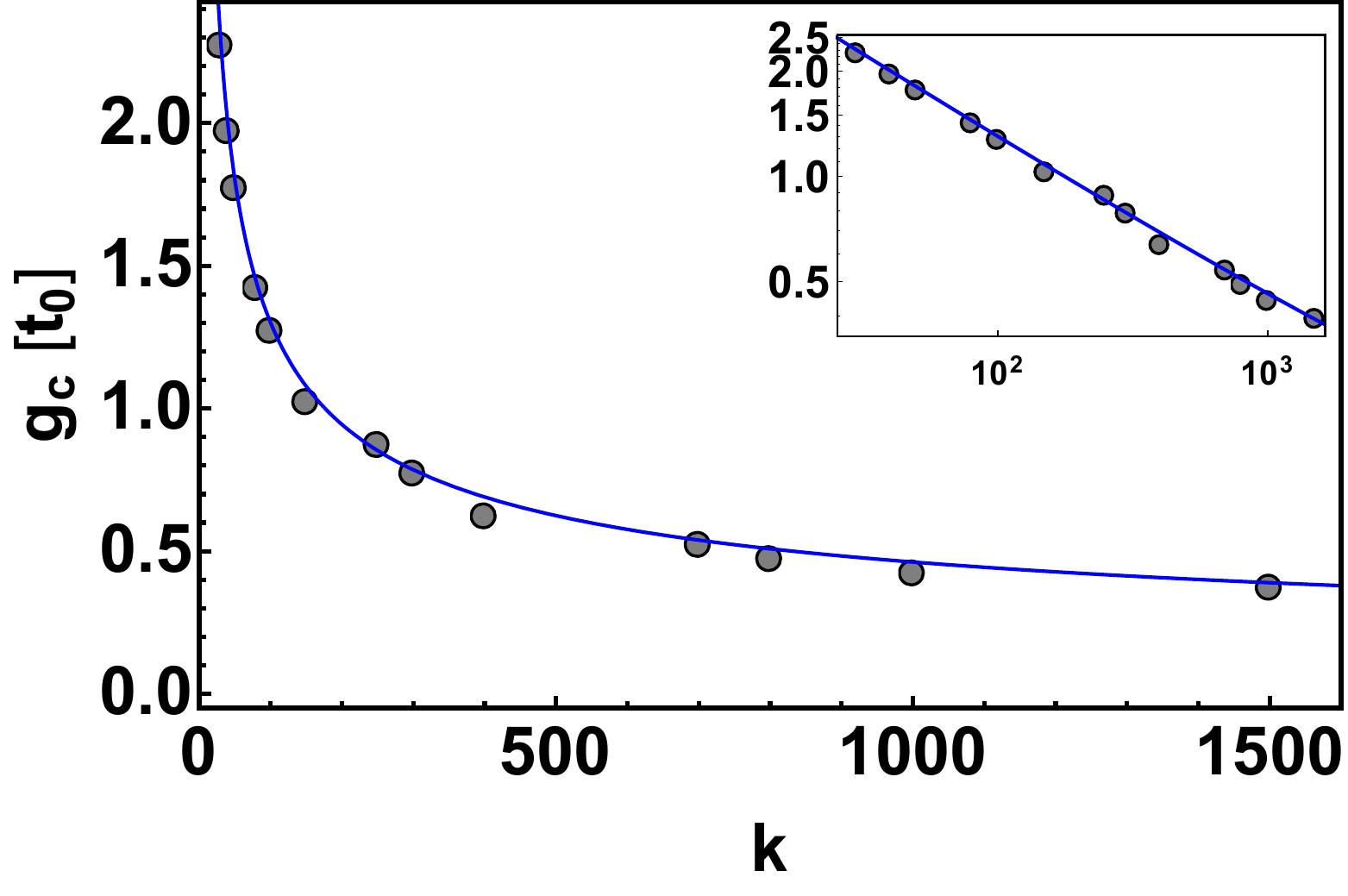}
\caption{
Critical electron-phonon coupling $g_c$ as a function of the scale $k$ (gray dots).
The continuum blue line has been obtained by fitting the data with the power law of Eq.~\eqref{fit}. 
The inset panel represents the data in log-log scale. This result shows how as we consider effectively larger systems  the transition happens at smaller critical values of $g$. Moreover, the dependence on system size displays a power law behavior with a critical exponent of $0.5$.
}
\lb{fig4}
\end{figure}

Figures \ref{fig3} and \ref{fig4} show the rippling transition for systems of different effective sizes, controlled by the scaling parameter $k$. For any value of $k$, as $g$ increases from zero there is a critical value, $g_c$, at which the flat solution loses its stability and the membrane displays rippling. This is shown in Fig.~\ref{fig3}(a) by the average local strain $\sqrt{\langle s^2\rangle}$ that acts as an order parameter: it is zero in the flat phase and positive in the rippled phase. Note that the phase transition is discontinuous. However, Figs.~\ref{fig3}(a) and (c) illustrate that the jump in the average local strain decreases as the scaling parameter increases. As $k$ increases, the critical value $g_c$ decreases. When plotted against $k$, as shown in figure \ref{fig4}, $g_c$ follows a power law with a critical exponent $\sim 0.5$. A simple fit using the model function:
\begin{equation}\label{fit}
g_c(k)=g_{c,\infty}+C\, k^{-\alpha}, 
\end{equation}
gives: $g_{c,\infty}=0.06t_0$, $C=11.8t_0$ and $\alpha=0.49$.
For large system sizes (large $k$), the small ratio $g_{c,\infty}/t_0$ makes the rippling configuration of freestanding  graphene stable even for very small couplings, which is consistent with the experimental observations \cite{meyer_nat07} of unavoidable rippling in suspended graphene monolayers. Moreover, we can rewrite equation \eqref{fit} as,
\begin{equation}
\frac{1}{k} \sim (g_c - g_{c,\infty})^2,
\label{para}
\end{equation}
this result suggests that there is a continuous line of bifurcations in the plane $(\frac{1}{k},g)$ (continuous line in Fig. \ref{fig5} (a)). Since this line cannot cross the axis $1/k=0$, its Taylor expansion around $g_{c,\infty}$ cannot have a linear term in $ (g_c - g_{c,\infty})$. Then, the transitions for different $k$ displayed in Fig. \ref{fig3} are capturing the second order term of the Taylor expansion, equation \eqref{para}. This explains the observed exponent $-1/2$ in equation \eqref{fit}.

 \begin{figure}
\includegraphics[width=0.45\textwidth]{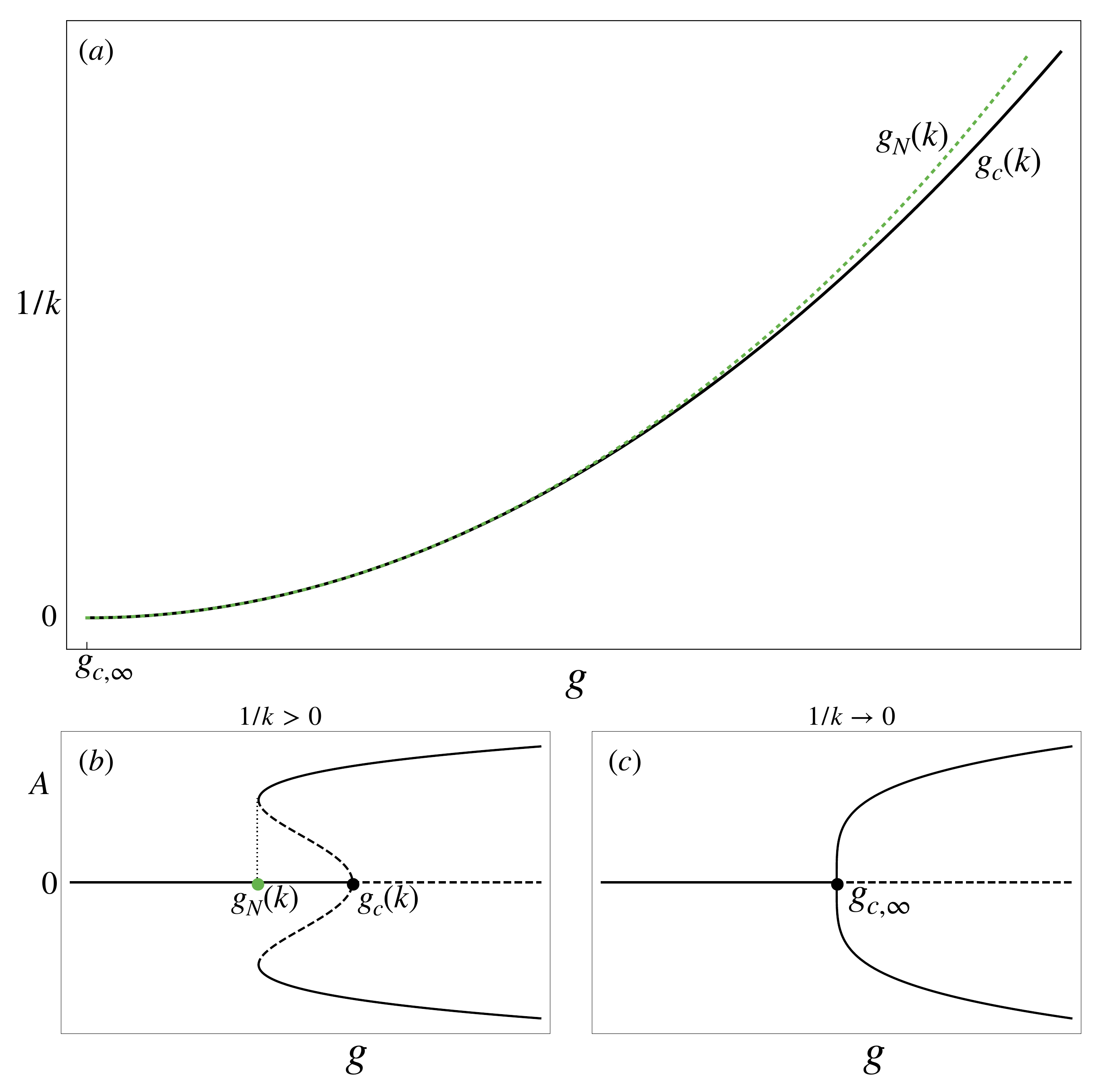}
\caption{Suggested bifurcation scheme for the flat and rippled configurations in the $(\frac{1}{k},g)$ plane. Panel (a) presents a continuous black line where the flat configuration becomes unstable.  For $g>g_c(k)$ the flat configuration is unstable, $g_N(k)<g<g_c(k)$ is a region of bistability, whereas, in the  $g<g_N(k)$ region, the flat configuration is the only solution. In panels (b) and (c), $A$ is the amplitude of the bifurcating rippled phase (order parameter), for example the averaged strain displayed in Fig. \ref{fig3}(a); $A=0$ is the flat configuration. Continuous and dashed lines correspond to stable and unstable solutions, respectively. For $1/k>0$ (panel (b)), the there is a subcritical bifurcation (first order transition), with a region of bistability $g_N(k)<g<g_c(k)$. Green and black points correspond to the crossing of the lines $g_N(k)$ and $g_c(k)$ in panel (a) for a specific value of $1/k$. The dotted line in panel (b) is a guide to the eye. In panel (c), for $\frac{1}{k} = 0$ the bifurcation has become supercritical (second order) and the bistability region has shrunk to zero}. 
\lb{fig5}
\end{figure}

For each value of $k$, the average strain of Fig.~\ref{fig3}(a) acts as an order parameter to characterize the amplitude of the ripples, analogous to $A$ in Fig. \ref{fig5} (b) and (c). Even though the order parameter undergoes a jump at the transition, see figure \ref{fig3} (a), the jump size decreases as $k$ increases.  Although a detailed numerical characterization of the order of the transition at every $k$ is left for future work, we propose the following plausible scenario: the bifurcation from the flat to rippled configuration is supercritical (second order transition) at infinite size $\frac{1}{k} = 0$, and is subcritical (first order) for $\frac{1}{k} > 0$, with a bistability region that increases its width as $\frac{1}{k}$ increases, see Fig. \ref{fig5}.  In Appendix \ref{app_3}, we  use the equation of a two-parameter pitchfork bifurcation with coefficients that are smooth in $1/k$ and $g$. We propose that there is an exceedingly sharp discontinuous transition with critical value given by Eq.~\eqref{fit} for large scaling parameter $k$. In this scenario, at $\frac{1}{k}=0$, the transition becomes continuous with a critical exponent $1/4$, such that $A\sim (g-g_{c,\infty})^{1/4}$. For $\frac{1}{k}>0$, there is a narrow bistability range of width $\propto k^{-2}$ ($g_N(k)<g<g_c(k)$). Since $k^{-2} \ll k^{-1/2}$ as $k\to\infty$, this bistability range is too narrow to be appreciable by our numerical iterations. Finally, this scenario also predicts that the size of the jumps should decrease as $k^{-\frac{1}{2}}$, in good agreement with Fig. \ref{fig3}(c).

\section{DISCUSSION}

\label{sec_dis}


We have studied an elastic graphene membrane coupled to its density of electrons. We iteratively solved the equations of elasticity, discretized on the honeycomb lattice, where the electronic density acts as source field. At the same time, the ground state for the electron density (which also depends on the membrane strain) is self-consistently determined within the tight-binding approximation. We find a critical value $g_c$ of the parameter controlling the coupling between deformations and electronic charge, above which a stable rippled phase appears. Upon scaling our equations to account for larger system sizes, $g_c$ decreases as a power law with critical exponent $\sim 1/2$, until it reaches a fixed value. We propose that the rippling transition of freestanding graphene is of second order in the limit of large system sizes ($k \to \infty$) whereas for finite system sizes the transition is of first order, with a bistability region and jump size that increase as $k$ decreases. A more detailed analysis of the bifurcation is left for future work. Moreover, we find a density of states (DOS) for the electronic eigenvalues that strongly depends on the coupling with the membrane deformations. The DOS suggests that a band gap opens up in the band structure as $g$ increases.

Although we have not systematically studied the size and structure of the ripples, we have checked that the typical ripple size decreases as $g$ increases deep inside the region of stable ripple configurations. Our numerical simulations suggest that the typical ripple size remains constant when the number of lattice points increases (for constant values of all the other parameters). Thus, the typical size of ripples seems to be independent of system size, although a more systematic study of this effect is left for future work. One effect we have not included in our numerical simulations is the long range Coulomb interaction. The Coulomb interaction suppresses charge accumulation at large length scales, and it is a marginal interaction when compared to the electronic kinetic energy. Thus, we do not expect its inclusion to change qualitatively our results. 
 
To our knowledge, this is the first time that this model has been exactly solved in the literature, and we have provided numerical support to previous analytical works~\cite{gazit_prb09,sanjose_prl11,guinea_prb14,bonilla2016critical}. But more importantly, we have shown how the problem of coupling membrane elasticity and electron density can be tackled numerically. In contrast with previous analytical work, we do not need to resort to any assumption for the electronic band structure: this is automatically taken care of by the tight-biding approach. Indeed, when the deformations are small, we recover the well-known Dirac cones in the density of states, whereas for larger ripples (as the coupling, $g$, increases) the band structure departs from the ideal case.

Our work paves the path to include the coupling to electron density on large scale molecular dynamics (MD) simulations of membrane mechanics. MD simulations have successfully predicted the mechanical behavior of graphene and other elastic membranes under thermal effects~\cite{xu2010geometry,bowick2017non}, whereas this work presents a numerical scheme to address the coupling between membrane elasticity and electronic density at zero temperature. Now, the interplay between temperature and phonon-electron coupling remains to be uncovered. The method presented in this work could be included as an intermediate step in MD simulations. This opens the possibility of studying the mechanics of $2$D materials in the most realistic framework, which includes thermal fluctuations and electron-phonon interactions.

\section*{Acknowledgements}
 
M.R-G acknowledges support from the National Science Foundation (USA) and
the Simons Foundation via awards 327939 and 454945. This work has been supported by the FEDER/Ministerio de Ciencia, Innovaci\'on y Universidades -- Agencia Estatal de Investigaci\'on grant MTM2017-84446-C2-2-R (LLB), and by funding from the European Commission under the Graphene Flagship, Core 3, grant number 881603 (TC and FG).
 
\appendix

\section{Discretizing the elasticity equations and the coupling to the electrons on the honeycomb lattice }
\label{app_2}
In the following we describe the procedure used to discretize the
continuum elasticity equations (10) of the main text on the honeycomb lattice,
which basically recalls the approach of Ref.s \cite{carpio_prb2008,carpio_prb2012}. We also explain the general method that we used to solve the problem of the elastic membrane coupled to the electrons.

\begin{figure}\centering
\includegraphics[scale=0.5]{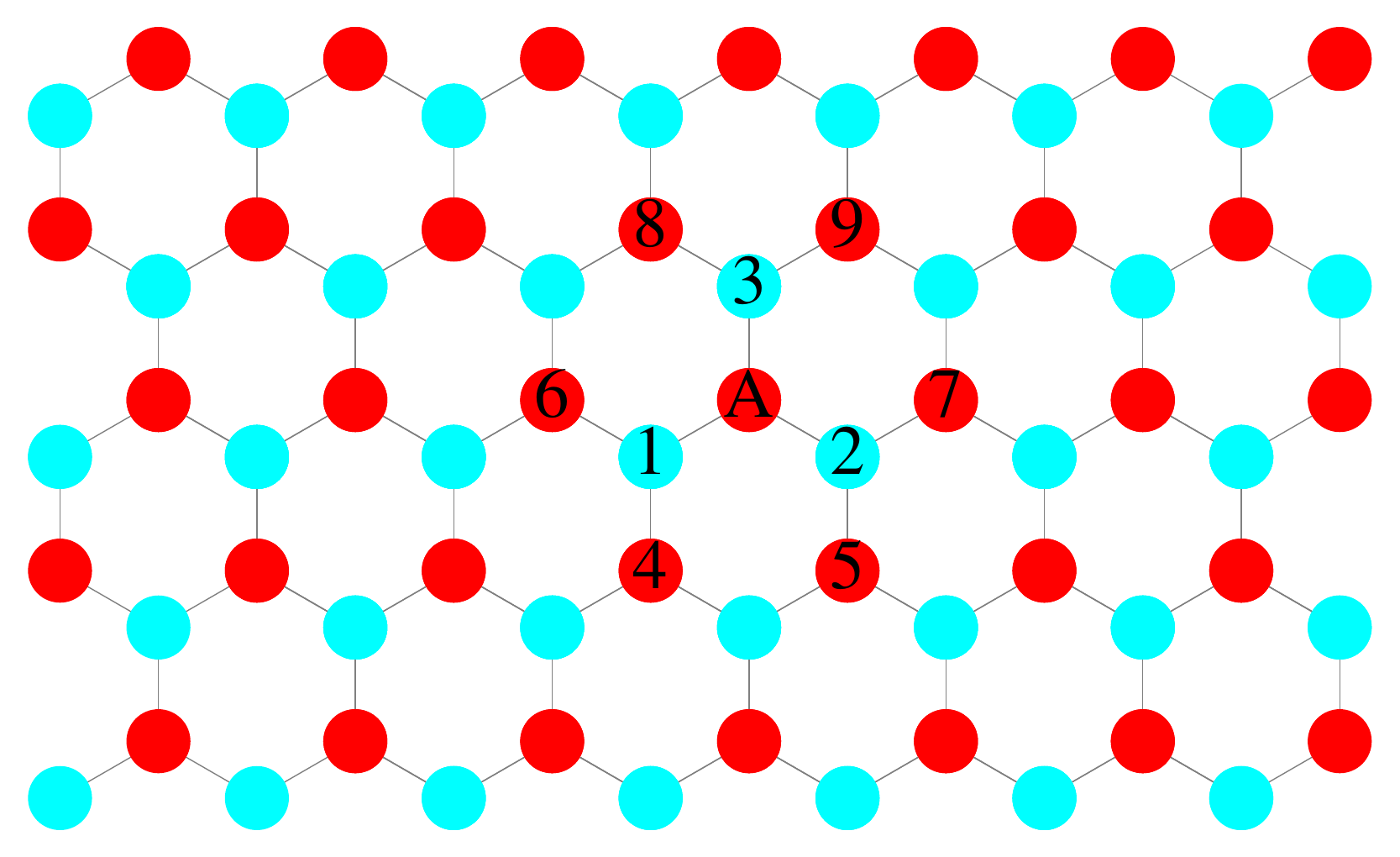}
\caption{
Neighbors of a given site of type $A$ (red). There are three first nearest neighbors $1,2,3$ of type $B$ (cyan) and six second nearest neighbors $4,\dots9$ of type $A$.
}\label{lattice}
\end{figure}

As usual, we describe the honeycomb lattice as consisting of two sub-lattices, that here we call of type $A$ and $B$.
This is depicted in the Fig. \ref{lattice}, where the atoms of type $A$ are represented in red and those of type $B$ in cyan,
although in the case of graphene the two species correspond to identical carbon atoms.
Each atom of type $A$ has three first nearest neighbors of type $B$,
that we labeled with the indices $1,2,3$, and six second nearest neighbors of type $A$,
labeled by the indices $4,\dots9$. If the atom $A$ has coordinates $(x,y)$ then, according to the scheme of the Fig. \ref{lattice}, the coordinates of its nine nearest neighbors are:
\bea
n_1=\left(x-\frac{a}{2},y-\frac{a}{2\sqrt{3}}\right)&,&n_2=\left(x+\frac{a}{2},y-\frac{a}{2\sqrt{3}}\right),\nn\\
n_3&=&\left(x,y+\frac{a}{\sqrt{3}}\right)\\
n_4=\left(x-\frac{a}{2},y-\frac{a \sqrt{3}}{2}\right)&,&n_5=\left(x+\frac{a}{2},y-\frac{a\sqrt{3}}{2}\right),\nn\\
n_6=\left(x-a,y\right)&,&n_7=\left(x+a,y\right),\nn\\
n_8=\left(x-\frac{a}{2},y+\frac{a\sqrt{3}}{2}\right)&,&n_9=\left(x+\frac{a}{2},y+\frac{a\sqrt{3}}{2}\right),\nn
\eea
where $a$ stands for the lattice constant.

 An analogous scheme, of course, holds for each atom of type $B$, with a similar definition of the nearest neighbors.

The first and second order partial derivatives that appear in the continuum equations (10) of the main text can be replaced by their corresponding finite differences on the lattice by introducing the following operators \cite{carpio_prb2008,carpio_prb2012}:
\begin{subequations}\lb{finite_diff}
\bea
Tf(A)&=&f(n_1)-f(A)+f(n_2)-f(A)\\
&+&f(n_3)-f(A)\sim \frac{a^2}{4}\nabla^2f\nn\\
Hf(A)&=&f(n_6)-f(A)+f(n_7)-f(A)\\
&\sim& a^2\partial^2_xf\nn\\
Df(A)&=&f(n_4)-f(n_5)+f(n_9)-f(n_8)\\
&\sim&a^2\sqrt{3}\partial_x\partial_yf.\nn\\
\Delta_xf(A)&=&f(n_2)-f(n_1)\sim a\partial_xf\lb{Dx}\\
\Delta_yf(A)&=&\frac{f(n_3)-f(A)-\left[f(n_1)-f(A)+f(n_2)-f(A)\right]}{2}\nn\\
&\sim&\frac{a}{\sqrt{3}}\partial_yf,\lb{Dy}\\
Bf(A)&=&Tf(n_1)-Tf(A)+Tf(n_2)-Tf(A)\nn\\&+&Tf(n_3)-Tf(A)\lb{B}
\eea
\end{subequations} 
where $f$ is a generic function of the lattice positions.

The continuum electronic density $\rho$ and its deviation $\d\rho$ appearing in the rhs of the Eq.s (10)
of the main text can be replaced by the occupation number per site $n(\mathbf{R})$
and by $\d n(\mathbf{R})=n(\mathbf{R})-n_0$,
$n_0$ being the filling (we use $n_0=1$).
Thus, the elasticity equations on the honeycomb lattice can be written as:
\begin{widetext}
\begin{subequations}\lb{latteleq}
\bea
4\mu Tu_x&+&(\lambda+\mu) Hu_x+\frac{\lambda+\mu}{\sqrt{3}}Du_y+\frac{\lambda+\mu}{a}\left[	\Delta_xhHh+\Delta_yh Dh	\right]+\frac{4\mu}{a}\Delta_xhTh=g\frac{\Delta_x \delta n}{a},
\lb{lattelequx} \\
4\left(\lambda+2\mu\right)Tu_y&-&\left(\lambda+\mu\right)Hu_y+\frac{\lambda+\mu}{\sqrt{3}}Du_x+\frac{4\sqrt{3}}{a}\left(\lambda+2\mu\right)\Delta_yhTh+\frac{\lambda+\mu}{a\sqrt{3}}\left[
\Delta_xhDh-3\Delta_yhHh\right]=\nn\\
&=&g\frac{\sqrt{3}\Delta_y \delta n}{a},\\
\frac{(\lambda+2\mu)}{a}\left\{\Delta_xh\left(
Hu_x\right.\right.&+&\left.\left.2\Delta_yhDh/a+\Delta_xhHh/a\right)
+\Delta_yh\left[\sqrt{3}\left(4T-H\right)u_y	+3\Delta_yh\left(4T-H\right)h/a\right]\right\}+\nn\\
+\frac{(\lambda+\mu)}{a}\left[
Du_x\Delta_yh\right.&+&\left.\Delta_xhDu_y/\sqrt{3}
\right]+
\frac{4\lambda Th}{a}\left[	\Delta_xu_x+\sqrt{3}\Delta_yu_y	\right]+\frac{2(\lambda+2\mu)}{a^2}Th\left[\left(\Delta_xh\right)^2+3\left(\Delta_yh\right)^2\right]+
\nn\\
+\frac{\mu}{a}\left[\sqrt{3}Hu_y\Delta_yh\right.&+&\left.	\left(4T-H\right)u_x\Delta_xh+2\Delta_xu_xHh+2\sqrt{3}\Delta_yu_y\left(4T-H\right)h+2Dh\left(\Delta_yu_x+\Delta_xu_y/\sqrt{3}\right)	\right]-\nn\\
-\frac{16\kappa}{a^2}Bh&=&\frac{g}{a^2}\left[
4Th\delta n+\Delta_xh\Delta_x\delta n+3\Delta_yh\Delta_y\delta n
\right].
\lb{lattelequy}
\eea
\end{subequations}
\end{widetext}

\section{Derivation of the F\"{o}ppl-von K\'arm\'an equations}
\label{app_1}
Here we show how to use the Airy potential to rewrite elasticity equations \pref{eq_cond2} of the main text in the form of F\"{o}ppl-von K\'arm\'an equations.

First of all, it is useful to introduce the stress tensor
\bea\lb{stress}
\s_{ij}=\l u_{kk}\d_{ij}+2\m u_{ij},
\eea
which allows us to write Eqs. \pref{eq_cond2} as:
\begin{subequations}
\bea
\partial_i\left[\s_{ij}\right.&-&\left.g\left(\d\rho\right)\d_{ij}\right]=0 \qquad i,j=x,y\lb{sigma_eq1}\\
\partial_i\left\{\left[\s_{ij}\right.\right.&-&\left.\left.g\left(\d\rho\right)\d_{ij}\right]\partial_jh\right\}-\kappa \left(\nabla^2\right)^2h=0.\lb{sigma_eq2}
\eea
\end{subequations}
Eq. \pref{sigma_eq1} is identically satisfied if we define the Airy potential, $\chi$, as:
\bea\lb{Airy_def}
&\s_{xx}-g\d\rho=\partial^2_y\chi,\\ &\s_{yy}-g\d\rho=\partial^2_x\chi,\\ &\s_{xy}=-\partial_x\partial_y\chi,
\eea
whereas Eq. \pref{sigma_eq2} becomes:
\bea\lb{chi_eq}
\kappa \nabla^2h-2[\chi,h]=0,
\eea
where 
\bea
[\chi,h]\equiv \frac{1}{2}\left[  \partial^2_x \chi\partial^2_y h+  \partial^2_y \chi\partial^2_x h-2\left(\partial_x\partial_y\chi\right)\left(\partial_x\partial_yh\right) \right].
\eea
To get an equation for $\chi$, we first check that
\bea\lb{curvature}
2\partial_x\partial_yu_{xy}-\partial^2_xu_{yy}-\partial^2_yu_{xx}=[h,h].
\eea
Then we use Eqs. \pref{stress} and \pref{Airy_def} to rewrite the left hand side of Eq. \pref{curvature} in terms of the Airy potential,
\bea
\frac{1}{Y}\nabla^2\chi+[h,h]=-\frac{g}{2B}\nabla^2\d\rho,
\eea
where $Y=\frac{4\mu(\l+\m)}{\l+2\m}$ and $B=\l+\m$
are the Young and compression moduli, respectively.
This equation, along with Eq. \pref{chi_eq}, represent the F\"{o}ppl-von K\'arm\'an equations for the membrane.

\section{Crossover argument from bifurcation theory}
\label{app_3}
Our numerical simulations suggest that the rippling transition becomes continuous as the size scaling $k\to\infty$. How can we understand this from bifurcation theory?

Let us assume that the rippling transition is a pitchfork bifurcation that goes from subcritical (discontinuous) to supercritical (continuous) as $k\to\infty$. Suppose that $A$, given by some average of $h$, characterizes the amplitude of the ripple state that bifurcates from the flat configuration. For fixed $1/k$, the bifurcation equation for $A$ is 
\begin{equation}
A\!\left[\tau+2\eta\!\left(\frac{1}{k}\right)\! A^2-A^4\right]\!=0, \quad \tau=g-g_c(k), \label{bifur}
\end{equation}
with $\eta(1/k)>0$. The bifurcation equation is written here as an expansion in powers of $A$ that is invariant under the transformation $A\to -A$. Terms of order $A^7$ and higher can be ignored near the bifurcation point and the sharpness of the numerically observed transition suggests that $\eta(0)=0$.  We have redefined $A$ so that the coefficient of $A^5$ is $-1$. The nonzero solutions of this equation satisfy
\begin{equation}
A_\pm^2= \eta(1/k)\pm \sqrt{\eta(1/k)^2+\tau}.  \label{bifur_sols}
\end{equation}\\
Eq.~\eqref{fit} with exponent $1/2$ can be reinterpreted as $1/k=(g_c-g_{c,\infty})^2/C^2$, which suggests that the line of critical values is close to a parabola in the plane $(g_c,1/k)$ with minimum at $1/k=0$. Assuming that the function $\eta(1/k)$ is smooth in $1/k$, $\eta(1/k)= \eta_0/k$ up to terms of order $1/k^2$.

A positive amplitude $A$ bifurcates subcritically from the flat configuration $A=0$ at $\tau=0$, it continues as $A_-$ for $\tau<0$ until $\tau=-\eta(1/k)^2=-\eta_0^2/k^2$, which is a turning point corresponding to $g_N(k)=g_c(k)-\eta_0^2/k^2$, and then it increases with $\tau$ as the branch $A_+$. Changing $A\to -A$, we find the corresponding solution with negative amplitude. Typically and as indicated in fig.~\ref{fig5}(b), $A=0$ is linearly stable for $\tau<0$, i.e., $g<g_c(k)$. The subcritical branch $A_-$, which exists for $g_N(k)\leq g\leq g_c(k)$, is unstable, whereas the higher amplitude branch $A_+$ exists for $g\geq g_N(k)$ and is stable. For fixed $k$, the discontinuous transition occurs at some $-\eta_0^2/k^2<\tau_1(k)=-r\eta_0^2/k^2<0$ ($0<r<1$), i.e., $g_N(k)<g_1(k)<g_c(k)$, for which the basins of attraction of $A_+$ and $A=0$ have the same size. As $k\to\infty$, both the turning point $g_N(k)$ and $g_1(k)$ collapse to the bifurcation point $g_c(\infty)=g_{c,\infty}$, and the discontinuous transition becomes a rather flat continuous transition with $A=\tau^{1/4}=(g-g_{c,\infty})^{1/4}$ for $g>g_{c,\infty}$, cf Fig.~\ref{fig5}(c). From Eqs.~\eqref{fit} and \eqref{bifur}, the discontinuous transition occurs at $g_1(k)=g_{c,\infty}+C k^{-1/2}-r\eta_0^2/k^2$. Since $k^{-2}\ll k^{-1/2}$ as $k\to\infty$, the last term is negligible compared to the others, and we also have $g_1(k)\sim g_c(k)$. At the turning point, Eq.~\eqref{bifur_sols} implies that the jump in amplitude is $[A]=\sqrt{\eta_0/k}$. 

In conclusion, together with our numerical simulations, bifurcation theory suggests that (i) the critical exponent is $1/4$ at $k=\infty$, $A\sim (g-g_{c,\infty})^{1/4}$, and (ii) the crossover from $A=0$ (flat membrane) to $A=A_+$ (rippling) occurs at a slightly smaller value than that given by Eq.~\eqref{fit}, $g_1(k)=g_c(k)-r\eta_0^2/k^2 \sim g_c(k)=g_{c,\infty}+Ck^{-1/2}$ as $k\to \infty$. The jump in amplitude at the turning point, which is also indicative of the discontinuous jump at $g=g_1(k)$, shrinks to zero as $k^{-1/2}$ as $k\to\infty$.

\bibliography{Literature.bib}



\end{document}